\documentclass[twocolumn,showpacs,preprintnumbers,amsmath,amssymb]{revtex4}
\usepackage{graphicx}
\usepackage{dcolumn}
\usepackage{bm}

\begin{document}
\title{How input fluctuations reshape the dynamics of a biological switching system}
\author{Bo Hu$^{1}$}
\author{David A. Kessler$^{2}$}
\author{Wouter-Jan Rappel$^{3}$}
\author{Herbert Levine$^{4}$}
\affiliation{$^{1}$ IBM T.J. Watson Research Center, PO Box 218, Yorktown Heights, NY, 10598, USA \\
$^{2}$ Department of Physics, Bar-Ilan University, Ramat-Gan, Israel \\
$^{3}$ Center for Theoretical Biological Physics, University of California San Diego La Jolla, CA, 92093-0319, USA \\
$^{4}$ Center for Theoretical Biological Physics, Rice University, Houston, TX, 77005, USA
}

\date{\today}
\begin{abstract}

An important task in quantitative biology is to understand the role of
stochasticity in biochemical regulation. Here, as an extension of our
recent work [Phys. Rev. Lett. \textbf{107}, 148101 (2011)], we study
how input fluctuations affect the stochastic dynamics of a simple
biological switch. In our model, the on transition rate of the switch
is directly regulated by a noisy input signal, which is described as a
nonnegative mean-reverting diffusion process. This continuous process can
be a good approximation of the discrete birth-death process and is
much more analytically tractable. Within this new setup, we apply the
Feynman-Kac theorem to investigate the statistical features of the output
switching dynamics. Consistent with our previous findings, the input
noise is found to effectively suppress the input-dependent transitions.
We show analytically that this effect becomes significant
when the input signal fluctuates greatly in amplitude and reverts slowly
to its mean.

\end{abstract}

\pacs{02.50.Le, 05.65.+b, 87.23.Ge, 87.23.Kg} \maketitle

\section{Introduction}

Stochasticity appears to be a hallmark of many biological processes involved in signal transduction and gene regulation \cite{Rao2002,Elowitz2002,Swain2002,Blake2003,Kaern2005,Raser2005,Pedraza2005,Cai2006,Friedman2006,Choi2008,Eldar2010}. Over the past decade, there have been numerous experimental and theoretical efforts to understand the functional roles of noise in various living systems \cite{Elowitz2002,Swain2002,Blake2003,Kaern2005,Raser2005,Pedraza2005,Cai2006,Friedman2006,Choi2008,Eldar2010,Paulsson2004,
Simpson2004,Bialek2005PNAS,Bialek2008PRL,Bialek2008PLoS,Bialek2009PRE,Shibata2005,tenWolde2006,tenWolde2009,LevineHwa2007,Wouter2008PRE,
Wouter2008PRL,Wouter2008PNAS,Fuller2010PNAS,Endres2008PNAS,Endres2009PRL,Bostani2012PRE,Hu2011PRL,Hu2011PRE,Hu2011JSP,Hu2010PRE,Hu2010PRL,
Cluzel2000,Cluzel2004,Cluzel2006,Emonet2008,Tu2005,Tu2008,Tu2010,Hornos2005PRE}.
Remarkably, the building block of different regulatory programs is
often a simple two-state switch under the regulation of some noisy
input signal. For example, a gene network is composed of many
interacting genes, each of which is a single switch regulated by
specific transcription factors. At a synapse, switching of
ligand-gated ion channels are responsible for converting the
presynaptic chemical message into postsynaptic electrical signals,
while the opening and closing of each channel depends on the binding
of certain ligands (such as a neurotransmitter). In bacterial
chemotaxis, the cellular motion is powered by multiple flagellar
motors and each motor rotates clockwise or counterclockwise depending
on the level of specific response regulator (e.g., CheY-P in \emph{E.
coli}).  Irrespective of the context, the input signal, i.e., the
number of regulatory molecules, is usually stochastic due to
discreteness, diffusion, random birth and death. How does a biological
switching system work in a noisy environment? This is the central
question we attempt to address in this paper.

Previous studies on this topic have usually focused on the
approximate, static relationship between input and output variations
(e.g., the \emph{additive noise rule})
\cite{Paulsson2004,Simpson2004,Bialek2005PNAS,Bialek2008PRL,Bialek2008PLoS,Bialek2009PRE,Shibata2005,
tenWolde2006}, while the dynamic details (e.g., dwell time statistics)
of the switching system have often been ignored. Our recent work
suggests that there is more to comprehend even in the simplest
switching system \cite{Hu2011PRL}.
For example, we showed that
increasing input noise does not always lead to an increase in the
output variation, disagreeing with the \emph{additive noise rule} as
derived from the coarse-grained Langevin approach. Traditional methods
often use a single Langevin equation to approximate the joint
input-output process, and relies on the assumption that the input
noise is small enough such that one can linearize input-dependent
nonlinear reaction rates. Our approach to this problem is quite
different as we explicitly model how the input stochastic process
drives the output switch, without making any small noise assumption.

In our previous paper \cite{Hu2011PRL}, the input signal was
generated from a discrete birth-death process and regulated the on transition
of a downstream switch. By explicitly solving the joint master
equation of the system, we found that input fluctuations can
effectively reduce the on rate of the switch. In this paper, this
problem is revisited in a continuous noise formulation. We propose to
model the input signal as a general diffusion process, which is
mean-reverting, nonnegative, and tunable in Fano factor (the ratio of
variance to mean). We employ the Feynman-Kac theorem to calculate the
input-dependent dwell time distribution and examine its asymptotic
behavior in different scenarios. Within this new framework, we recover
several  findings reported in  \cite{Hu2011PRL}, and also demonstrate
how the noise-induced suppression depends on the relative noise level
as well as the relative input-to-output timescale. Finally, we
elaborate on how the diffusion process introduced in this paper can be
a reasonable approximation of the discrete birth-death process.

\section{Model}
The input of our model, denoted by $X(t)$, represents a specific chemical
concentration at time $t$ and directly governs the transition rates of a
downstream switch. The binary on-off states of the switch in continuous time
constitute the output process, $Y(t)$. A popular choice for $X(t)$ is the
Ornstein-Uhlenbeck (OU) process due to its analytical simplicity and
mean-reverting property \cite{Gardiner,vanKampen2007,Oksendal2000}. However, this
process does not rule out negative values, an unphysical feature for modeling
chemical concentrations. For both mathematical convenience and biophysical
constraints (see Section IV for more details), we model $X(t)$ by a square-root
diffusion process \cite{CIR1985,Pechenik1999PRE,Shibata2003PRE,Maritan2009PRE}:
\begin{equation}
dX(t)=\lambda[\mu-X(t)]dt+\sigma\sqrt{X(t)}dW_t,
\label{cir}
\end{equation} where $\lambda$ represents the rate at which $X(t)$
reverts to its mean level $\mu$, $\sigma$ controls the noise
intensity, and $W_t$ denotes the standard Brownian motion. This simple
process is known as the Cox-Ingersoll-Ross (CIR) model of
interest rates \cite{CIR1985}. The square-root noise term not only
ensures that the process never goes negative but also captures a common
statistical feature underlying many biochemical processes, that is,
the standard deviation of the copy number of molecules scales as the
square root of the copy number, as dictated by the \emph{Central Limit
Theorem}. Solving the Fokker-Planck equation for Eq. (1), we obtain
the steady-state distribution of the input signal:
\begin{equation}
P_s(X=x)=\frac{\beta^\alpha x^{\alpha-1}e^{-\beta x}}{\Gamma(\alpha)},
\ \ \ \alpha\equiv\frac{2\mu\lambda}{\sigma^2}, \ \ \ \beta\equiv\frac{2\lambda}{\sigma^2},
\end{equation}
which is a Gamma distribution with stationary variance $\sigma_X^2=\mu\sigma^2/(2\lambda)$. This is another attractive aspect of this model, since the protein abundance from gene expression experiments can often be fitted by a Gamma distribution \cite{Cai2006,Friedman2006,Choi2008}. The parameter $\alpha$ in Eq. (2) can be interpreted as the \emph{signal-to-noise ratio}, since $\alpha=\mu^2/\sigma_X^2$. For $\alpha\geq1$, the zero point is guaranteed to be inaccessible for $X(t)$. The other shape parameter $\beta$ sets the Fano factor as we have $1/\beta=\sigma_X^2/\mu$. Using the It\'{o} calculus \cite{Oksendal2000}, one can also find the steady-state covariance: $\lim_{{t\rightarrow\infty}}\mathrm{Cov}[X(t),X(t+s)]=\sigma_X^2 e^{-\lambda|s|}$. Thus $X(t)$ is a stationary process with correlation time $\lambda^{-1}$.

\begin{figure}
\scalebox{0.42}[0.42]{\includegraphics{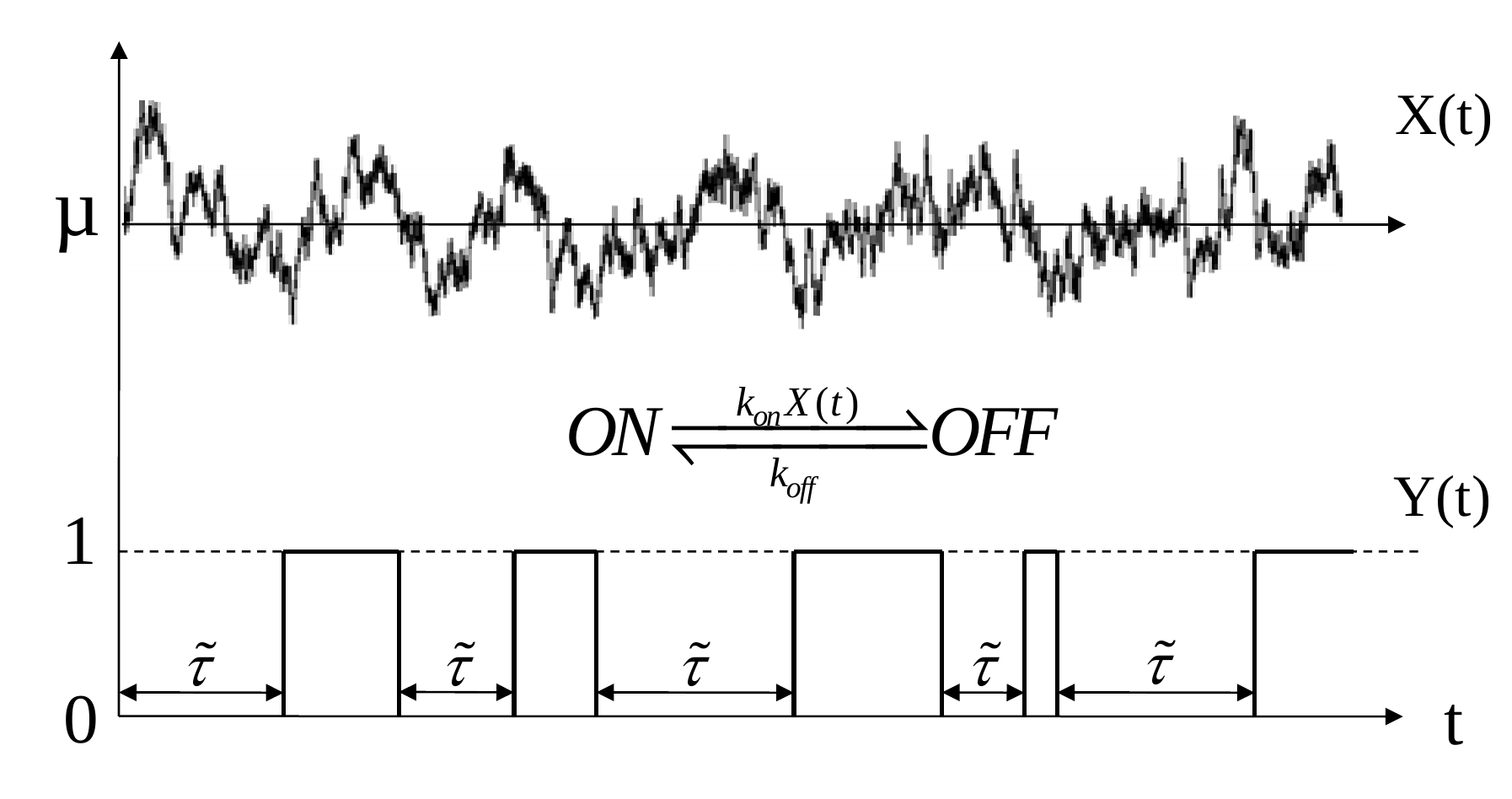}} \caption{Illustration of our model:
$X(t)$ represents the input signal which fluctuates around a mean level over time; $Y(t)$ records the switch process which flips between the off ($Y=0$) and on ($Y=1$) states with transition rates $k_{\mathrm{on}}X(t)$ and $k_{\mathrm{off}}$. $\widetilde{\tau}$ is the dwell time in the off state.}
\end{figure}

In reality, the output switching rates may depend on the input $X(t)$ in complicated ways, depending on the detailed molecular mechanism. For analytical convenience, we assume that the on and off transition rates of the switch are $k_{\mathrm{on}}X(t)$ and $k_{\mathrm{off}}$, respectively (Fig. 1). As a result, the input fluctuations should only affect the chance of the switch exiting the off state; the on-state dwell time distribution is always exponential with rate parameter $k_{\mathrm{off}}$. If we mute the input noise, then $Y(t)$ reduces to a two-state Markov process with transition rates $k_{\mathrm{on}}\mu$ and $k_{\mathrm{off}}$. However, the presence of input noise will generally make $Y(t)$ a non-Markovian process, because the off-state time intervals may exhibit a non-exponential distribution. To illustrate this point more rigorously, we consider the following first passage time problem \cite{Oksendal2000}. Suppose the switch starts in the off state at $t=0$ with the initial input $X(0)=x$. Let $\widetilde{\tau}$ be the first time of the switch turning on (Fig. 1). Then the \emph{survival probability} $f(x,t)$ for the switch staying off up to time $t$ is given by
\begin{equation}
f(x,t)\equiv P(\widetilde{\tau}>t|X(0)=x)=\mathrm{E}^{x}\left[e^{-\int_0^tk_{\mathrm{on}}X(s) ds}\right],
\end{equation}
where $\mathrm{E}^{x}[...]$ denotes expectation over all possible sample paths of $X(s)$ for $0\leq s\leq t$, conditioned on $X(0)=x$. The Feynman-Kac formula \cite{Oksendal2000} asserts that $f(x,t)$ must solve the following partial differential equation:
\begin{equation}
\frac{\partial f}{\partial t}=\lambda(\mu-x)\frac{\partial f}{\partial x}
+\frac{1}{2}\sigma^2 x\frac{\partial^2 f}{\partial x^2}-k_{\mathrm{on}}xf,
\label{pdef}
\end{equation}
with the initial condition $f(x,0)=1$. As we will show in the next section, the off-state dwell time distribution is not exactly exponential, though it is asymptotically exponential when $t$ is large.

\section{Results}
A similar partial differential equation to Eq. (\ref{pdef}) has been solved in Ref. \cite{CIR1985} to price zero-coupon bonds under the CIR interest rate model. The closed-form solution for our problem is similar and is found to be:
\begin{equation}
f(x,t)=\left[\frac{\widetilde{\lambda}e^{\lambda t/2}/\sinh(\widetilde{\lambda}t/2)}
{\lambda+\widetilde{\lambda}\coth(\widetilde{\lambda}t/2)}\right]^{\alpha}
\exp\left[\frac{-2k_{\mathrm{on}}x}{\lambda+\widetilde{\lambda}\coth(\widetilde{\lambda}t/2)}\right]
\label{fxt}
\end{equation}
where
\begin{equation}
\widetilde{\lambda}\equiv\sqrt{\lambda^2+2k_{\mathrm{on}}\sigma^2} =\lambda\sqrt{1+2k_{\mathrm{on}}\sigma^2/\lambda^2}.
\end{equation}
Evidently, $f(x,t)$ remembers the initial input $x$ and decays with
$t$ in a manner which is not exactly exponential. For $t \gg
\widetilde{\lambda}^{-1}$, Eq. (\ref{fxt}) takes the following form,
\begin{equation}
f(x,t)\simeq
\left(\frac{2\widetilde{\lambda}}{\lambda+\widetilde{\lambda}}\right)^{\alpha}
\exp\left(-\frac{\widetilde{k}_{\mathrm{on}}x}{\lambda}-\widetilde{k}_{\mathrm{on}}\mu t\right).
\label{asypf}
\end{equation}
In deriving Eq. (\ref{asypf}), we have used the following relationship which defines the new parameter $\widetilde{k}_{\mathrm{on}}$ as below:
\begin{equation}
\widetilde{k}_{\mathrm{on}}\equiv \frac{\alpha(\widetilde{\lambda}-\lambda)}{2\mu}=\frac{\lambda}{\sigma^2}(\widetilde{\lambda}-\lambda)
=\frac{2\lambda}{\widetilde{\lambda}+\lambda}k_{\mathrm{on}}< k_{\mathrm{on}}.
\end{equation}
Thus, asymptotically speaking, $f(x,t)$ decays with $t$ in an exponential manner at the rate $\widetilde{k}_{\mathrm{on}}\mu$. It can be easily verified that Eq. (\ref{asypf}) is a particular solution to Eq. (\ref{pdef}).

To gain further insight into how the input noise affects the switching dynamics, we first study the ``slow switch" limit where $X(t)$ fluctuates so rapidly that $\lambda^{-1}\ll\mathcal{T}_Y$. Here, $\mathcal{T}_Y\equiv(k_{\mathrm{on}}\mu+k_{\mathrm{off}})^{-1}$ is the output correlation time for the noiseless input model. In this limit, the initial input $x$ in $f(x,t)$ at the start of each off period is effectively drawn from the Gamma distribution $P_s(x)$ defined in Eq. (2); the successive off-state intervals are almost independent with each other (as the input loses memory quickly) and are distributed as
\begin{equation}
P(\widetilde{\tau}\leq t)=1-P(\widetilde{\tau}>t)=1-\int_0^{\infty}f(x,t)P_s(x)dx.
\end{equation}
By direct integration over $x$, we find that
\begin{eqnarray}
P(\widetilde{\tau}>t)&=&\left[\frac{\beta\widetilde{\lambda}e^{\lambda t/2}}{(\beta\lambda+2k_{\mathrm{on}})\sinh(\widetilde{\lambda}t/2)
+\beta\widetilde{\lambda}\cosh(\widetilde{\lambda}t/2)}\right]^{\alpha} \nonumber\\
&\simeq&\left[\frac{2\beta\widetilde{\lambda}e^{-(\widetilde{\lambda}-\lambda)t/2}}
{\beta(\lambda+\widetilde{\lambda})+2k_{\mathrm{on}}}\right]^{\alpha}\simeq\exp(-\widetilde{k}_{\mathrm{on}}\mu t).
\end{eqnarray}
In the last step above, we have used Eq. (8) as well as the following observation: By introducing $\theta\equiv k_{\mathrm{on}}\sigma^2/\lambda^2$ which reflects the deviation of $\widetilde{\lambda}$ from $\lambda$, one can check that as long as
$\lambda\gg k_{\mathrm{on}}\mu$ (as ensured by $\lambda^{-1}\ll\mathcal{T}_Y$) the following holds, regardless of the values of $\theta$,
\[
\left[\frac{2\beta\widetilde{\lambda}}{\beta(\lambda+\widetilde{\lambda})+2k_{\mathrm{on}}}\right]^{\alpha}
=\left(\frac{1}{2}+\frac{1+\theta}{2\sqrt{1+2\theta}}\right)^{-\frac{2}{\theta}\cdot\frac{k_{\mathrm{on}}\mu}{\lambda}}\simeq1.
\]
Our simulations show that the approximate result, $P(\widetilde{\tau}>t)\simeq e^{-\widetilde{k}_{\mathrm{on}}\mu t}$, is excellent (Fig. 2A), independent of the values of $\theta$. Thus, the average waiting time for the switch to turn on is $(\widetilde{k}_{\mathrm{on}}\mu)^{-1}$, longer than the corresponding average time $(k_{\mathrm{on}}\mu)^{-1}$ for the noiseless input model. This is similar to our previous result \cite{Hu2011PRL}, and suggests that the input noise will effectively suppress the on state by increasing the average waiting time to exit the off state. Consequently, the probability, $P_{\mathrm{on}}$, to find the switch on ($Y=1$) is less than that in the noiseless input model:
\begin{equation}
P_{\mathrm{on}}\simeq\frac{\mu}{\mu+\widetilde{K}_d}<\frac{\mu}{\mu+K_d}=\lim_{\sigma\rightarrow0}P_{\mathrm{on}},
\end{equation}
where $\widetilde{K}_d\equiv
k_{\mathrm{off}}/\widetilde{k}_{\mathrm{on}}$, the \emph{effective}
equilibrium constant, is larger than the original $K_d\equiv
k_{\mathrm{off}}/k_{\mathrm{on}}$ as per Eq. (8).

\begin{figure}
\scalebox{0.45}[0.45]{\includegraphics{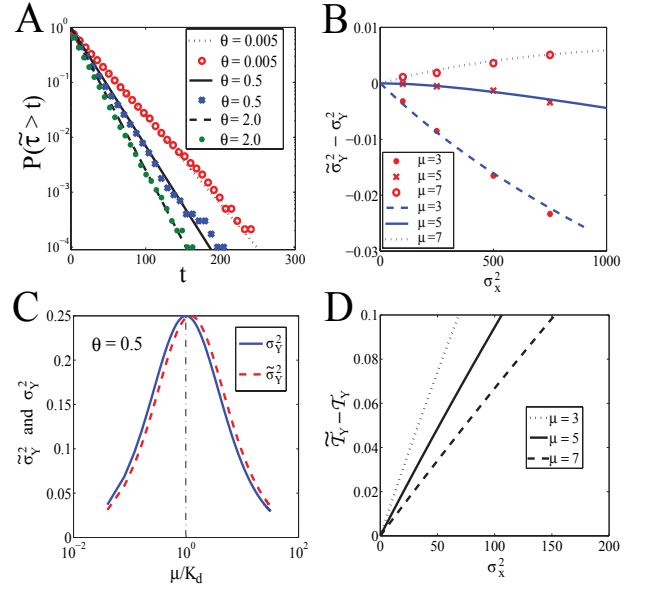}} \caption{(color online). The slow switch case. Here we use
$\lambda=10$, $k_{\mathrm{on}}=0.02$, and $k_{\mathrm{off}}=0.1$ (thus $K_d=5$).
(A) $P(\widetilde{\tau}>t)$ versus $t$ for $\mu=3$ and $\theta=0.005$, $0.5$, and $2.0$
which are achieved by choosing $\sigma=5$, $50$, and $100$. Symbols represent simulation results,
while lines denote $\exp(-\widetilde{k}_{\mathrm{on}}\mu t)$.
(B) $\widetilde{\sigma}_Y^2-\sigma_Y^2$ versus $\sigma_X^2$, where different values
of $\sigma_X^2$  are obtained by tuning $\sigma$. (C) $\widetilde{\sigma}_Y^2$ and $\sigma_Y^2$
versus $\mu/K_d$ with $\theta=0.50$. (D) $\widetilde{\mathcal{T}}_Y-\mathcal{T}_Y$ versus $\sigma_X^2$. }
\end{figure}

Therefore, in the slow switch limit ($\mathcal{T}_Y\gg\lambda^{-1}$), the output $Y(t)$ is approximately a two-state Markov process with transition rates $\widetilde{k}_{\mathrm{on}}\mu$ and $k_{\mathrm{off}}$. If we further assume that the noise is modest (i.e., $\sigma_X\leq\mu$), then
\begin{equation}
\theta\equiv k_{\mathrm{on}}\frac{\sigma^2}{\lambda^2}=\frac{2k_{\mathrm{on}}\mu}
{\alpha\lambda}=\frac{2k_{\mathrm{on}}\mu}{\lambda}\left(\frac{\sigma_X^2}{\mu^2}\right)\ll1.
\label{theta}
\end{equation}
The equality above shows that $\theta$ is a characteristic parameter determined by the ratio of the input to output time scales and the relative noise strength. For $\theta\ll1$, we have $\widetilde{\lambda}=\lambda\sqrt{1+2\theta}\simeq\lambda(1+\theta)$ by Eq. (6), and the effective on rate defined in Eq. (8) becomes,
\begin{equation}
\widetilde{k}_{\mathrm{on}}=\frac{2k_{\mathrm{on}}}{1+\sqrt{1+2\theta}}\simeq\frac{k_{\mathrm{on}}}{1+\frac{\theta}{2}}
=k_{\mathrm{on}}\left(1+\frac{k_{\mathrm{on}}}{\lambda}\frac{\sigma_X^2}{\mu}\right)^{-1}.
\end{equation}
In \cite{Hu2011PRL}, the input signal was taken to be a Poisson
birth-death process for which the variance is equal to the mean
($\sigma_X^2=\mu$), and the time scale has been normalized by putting
the death rate equal to one (which amounts to setting $\lambda=1$ here).
These two constraints reduce Eq. (13) to $\widetilde{k}_{\mathrm{on}}\simeq
k_{\mathrm{on}}/(1+k_{\mathrm{on}})$, recovering the result we derived
in \cite{Hu2011PRL}.
The consistency indicates that our key findings are general,
independent of the specific model we choose. The continuous diffusion
model here, however, is more flexible as it allows the Fano factor to
differ from one, that is, $\sigma_X^2\neq\mu$.

For small $\theta$, the stationary variance of $Y(t)$ can be expanded as follows:
\begin{eqnarray}
\widetilde{\sigma}_Y^2&=&P_{\mathrm{on}}(1-P_{\mathrm{on}})
\simeq \sigma_Y^2+\frac{\mu K_d (\mu-K_d)}{2(\mu+K_d)^3}\theta+\mathcal{O}(\theta^2), \nonumber \\
&=&\sigma_Y^2+\frac{\mu-K_d}{(\mu+K_d)^3}\frac{k_{\mathrm{off}}}{\lambda}\sigma_X^2+\mathcal{O}(\sigma_X^4),
\label{vardiff}
\end{eqnarray}
where $\sigma_Y^2\equiv\mu K_d/(\mu+K_d)^2$ is the output variance of the noiseless input model. Eq. (\ref{vardiff}) indicates that the input noise $\sigma_X^2$ does not always contribute positively to the output variance $\widetilde{\sigma}_Y^2$. In fact, the contribution is negligible when $\mu$ is near $K_d$ and even negative for $\mu<K_d$ (Fig. 2B). The explanation is, as we argued before \cite{Hu2011PRL}, that a two-state switch at any moment is a Bernoulli random variable whose variance is always bounded by one quarter (Fig. 2C). Finally, with the effective on rate $\widetilde{k}_{\mathrm{on}}$, the correlation time of $Y(t)$ becomes $\widetilde{\mathcal{T}}_Y = (\widetilde{k}_{\mathrm{on}}\mu+k_{\mathrm{off}})^{-1}$, and can likewise be expanded as follows:
\begin{equation}
\widetilde{\mathcal{T}}_Y = \mathcal{T}_Y\left(1-\frac{\widetilde{\lambda}-\lambda}{\widetilde{\lambda}+\lambda}
\frac{\mu}{\mu+K_d}\right)^{-1}
\simeq\mathcal{T}_Y+\frac{1}{\lambda}\frac{\sigma_X^2}{(\mu+K_d)^2}.
\end{equation}
Thus, $\widetilde{\mathcal{T}}_Y$ \emph{weakly} increases with the input noise in this small noise limit (Fig. 2D).

\begin{figure}
\scalebox{0.42}[0.42]{\includegraphics{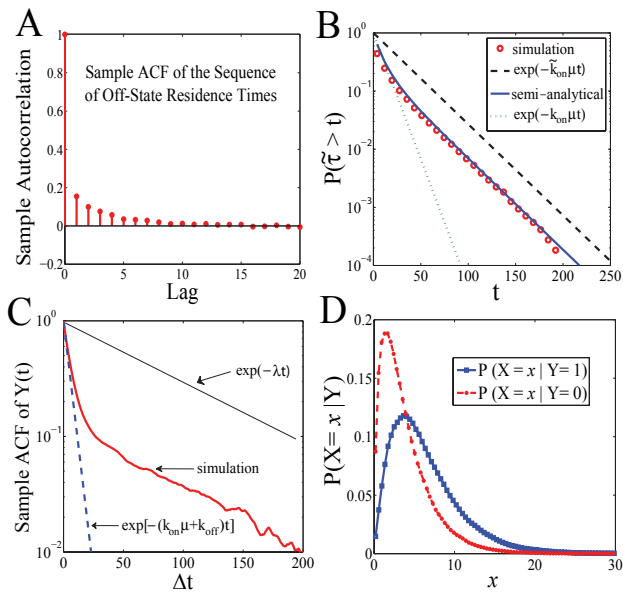}} \caption{(color online). The fast switch case. Here we
choose $\lambda=0.0125$, $\mu=5$, $k_{\mathrm{off}}=0.1$, $k_{\mathrm{on}}=0.02$,
and $\theta=10$ (such that $\sigma\simeq0.28$). (A) Sample ACF of successive
off-state time intervals. (B) Distribution of the off-state intervals, $P(\widetilde{\tau}>t)$.
Symbols are from Monte-Carlo simulations and solid line is the semi-analytical prediction described in the main text.
(C) Sample ACF of a simulated sample path of $Y(t)$ in the semi-log scale. (D) Input distributions conditioned on the output. }
\end{figure}

We now examine the ``fast switch" limit where the switch flips much faster than the input reverts to its mean ($\mathcal{T}_Y\ll\lambda^{-1}$). In this scenario, the initial input values $\{x_i, i=1,2,...\}$ for successive first-passage time intervals $\{\widetilde{\tau}_i, i=1,2,...\}$ are correlated due to the slow relaxation of $X(t)$. This memory makes the sequence $\{\widetilde{\tau}_i, i=1,2,...\}$ correlated as well, as confirmed by our Monte-Carlo simulations (Fig. 3A). For the same reason, the \emph{autocorrelation function} (ACF) of the output $Y(t)$ exhibits two exponential regimes (Fig. 3C): over short time scales, it is dominated by the intrinsic time $\mathcal{T}_Y$ of the switch; over long time scales, however, it decays exponentially at the input relaxation rate $\lambda$. This demonstrates that the long-term memory in $X(t)$ is inherited by the output process $Y(t)$. For a fast switch, the distribution $P(\widetilde{\tau}>t)$ is not fully exponential (Fig. 3B), though it decays exponentially at the rate of $\widetilde{k}_{\mathrm{on}}\mu$ for large $t$, as predicted by the asymptotic Eq. (7). We can still use the closed-form solution of $f(x,t)$ in Eq. (5) to fit the simulation data (open circles). Specifically, we calculate $P(\widetilde{\tau}>t)=\int_0^{\infty}f(x,t)P_{\widetilde{\tau}}(x)dx$ (blue line in Fig. 3B), where $P_{\widetilde{\tau}}(x)$ is the distribution of the initial $x$ for each switching event (defining the ``first-passage" time $\widetilde{\tau}$) and can be obtained from the same Monte-Carlo simulations. Clearly, such a semi-analytical trial (blue line) provides a nice fit to the simulation results (open circles).

Note that $P_{\widetilde{\tau}}(x)\neq P_s(x)$ due to memory effects in the fast switch limit. To show this, we illustrate the input-output interdependence by plotting the input distributions conditional on the output state in Fig. 3D. In fact, we shall have $P(X=x|Y=1)=P_{\widetilde{\tau}}(x)$, because: first, $x$ in $P_{\widetilde{\tau}}(x)$ denotes $x=X(t_0^+)$ where $t_0$ is the last time of off-transition; second, $X(t_0^+)=X(t_0^-)$ due to continuity and $Y(t_0^-)=1$ by definition; third, all the on-state intervals are memoryless with the same Poisson rate $k_{\mathrm{off}}$. This simple relation $P(X=x|Y=1)=P_{\widetilde{\tau}}(x)$ has been confirmed by our simulations (results not shown). It is also obvious from Fig. 3D that the expectation value of the input conditioned on $Y=1$ is larger than that given $Y=0$. Thus the mean of $X(t)$ should lie in between, i.e., $\mathrm{E}[X|Y=1]>\mathrm{E}[X]>\mathrm{E}[X|Y=0]$, an interesting feature of the fast switch limit \cite{Hu2011PRL}. Since $f(x,t)$ is a decreasing function of $x$ and the initial input $x$ is likely to be larger under $P_{\widetilde{\tau}}(x)=P(X=x|Y=1)$ than under the measure $P_{s}(X=x)$, we should have
\begin{eqnarray}
P(\widetilde{\tau}>t)&=&\int_0^{\infty}f(x,t)P_{\widetilde{\tau}}(x)dx \nonumber\\
&<&\int_0^{\infty}f(x,t)P_{s}(x)dx\simeq e^{-\widetilde{k}_{\mathrm{on}}\mu t}.
\end{eqnarray}
The above inequality explains why the distribution of
$\widetilde{\tau}$ is below the single exponential
$e^{-\widetilde{k}_{\mathrm{on}}\mu t}$ (dashed line) in Fig. 3B. All
the above results (Fig. 3A-D) indicate that the output process $Y(t)$
is non-Markovian in the fast switch limit and, again, confirm the more general applicability of our findings reported in \cite{Hu2011PRL}.

\section{Diffusion Approximation}

In this paper, we have used the square-root diffusion (or CIR) process
to model biochemical fluctuations. Here we argue that this choice is
inspired by the fundamental nature of general biochemical processes.
Many biochemical signals are subject to counteracting effects:
synthesis/degradation, activation/deactivation, transport in/out of
a cellular compartment, etc. As a result, these signals tend to fluctuate
around their equilibrium values. A simple yet realistic model to capture
these phenomena is the birth-death process, which we adopted to model
biochemical noise in \cite{Hu2011PRL}. Remarkably, a birth-death process
can be approximated by a Markov diffusion process \cite{Gardiner}.
The standard procedure is to employ the Kramers-Moyal expansion to convert
the master equation into a Fokker-Planck equation (if terminating
after the second term). This connection allows the use of a Langevin equation
to approximate the birth-death process. We will explain this in a more
intuitive way.

Assume that the birth and death rates for the input signal $X(t)$ are $\nu$ and $\lambda X(t)$. Then the stationary distribution of $X(t)$ is a Poisson distribution, with its mean, variance, and skewness given by $\mu\equiv\nu/\lambda$, $\sigma_X^2=\mu$, and $\mu^{-1/2}$, respectively. The corresponding Langevin equation that approximates this birth-death process is
\begin{equation}
\frac{dX(t)}{dt}=\nu - \lambda X(t) +\eta(t),
\label{lang}
\end{equation}
where the stochastic term $\eta(t)$ represents a white noise with $\langle\eta(t)\rangle=0$ and delta-correlation
\begin{equation}
\langle\eta(t)\eta(t')\rangle=(\nu+\lambda X)\delta(t-t')=\lambda(\mu+X)\delta(t-t').
\label{var}
\end{equation}
Physically, the Langevin approximation Eq. (\ref{lang}) holds when the copy number of molecules is large and the time scale of interest is longer than the characteristic time ($\lambda^{-1}$) of the birth-death process. Eq. (\ref{var}) indicates that the instantaneous variance of the noise term $\eta(t)$ equals the sum of the birth rate $\nu$ and the death rate $\lambda X(t)$. An intuitive interpretation is that since both the birth and death events follow independent Poisson processes, the total variance of the increment $X(t+\Delta t)-X(t)$ over a short time $\Delta t$ must be equal to $\nu\Delta t + \lambda X(t) \Delta t$. As $\mu\equiv\nu/\lambda$, we can rewrite the Langevin Eq. (\ref{lang}) as the following It\'{o}-type \emph{stochastic differential equation} (SDE):
\begin{equation}
dX(t)=\lambda[\mu-X(t)]dt+\sqrt{\lambda\mu+\lambda X(t)}dW_t,
\label{sde}
\end{equation}
which is similar to Eq. (1) introduced at the beginning. A transformation $X'(t)\equiv X(t)+\mu$ is convenient for exploiting our existing results, as the SDE for $X'(t)$ is
\begin{equation}
dX'=\lambda(2\mu-X')dt+\sqrt{\lambda X'}dW_t,
\label{sde2}
\end{equation}
which is a particular CIR process. It is easy to check that the stationary variance of $X'(t)$ equals $\mu$, and so does the variance of $X(t)$. In other words, we still have $\sigma_X^2=\mu$ for $X(t)$ under Eq. (\ref{sde}). As a CIR process, $X'$ in equilibrium follows a Gamma distribution, the skewness of which is found to be $\mu^{-1/2}$. Therefore, $X=X'-\mu$ follows a ``shifted" Gamma distribution with its mean, variance, and skewness given by $\mu$, $\mu$, and $\mu^{-1/2}$, respectively, the same to those of the Poisson distribution. This matching of moments suggests that Eq. (\ref{sde}) is indeed a nice approximation to the original birth-death process. However, $X(t)$ in Eq. (\ref{sde}) can take negative values, because $X=X'-\mu$ is bounded below by $-\mu$ due to its ``shift" in distribution. This becomes a limitation of the Langevin approximation Eq. (\ref{lang}) which could fail if the noise is large (i.e. the number of molecules is small). 

Under Eq. (\ref{sde}) for the input $X(t)$, we can still evaluate the analog of Eq. (3):
\begin{eqnarray}
f(x,t)&=&\mathrm{E}^{x}\left[e^{-\int_0^tk_{\mathrm{on}}X(s) ds}\right] \nonumber\\
&=&\mathrm{E}^{x+\mu}\left[e^{-\int_0^tk_{\mathrm{on}}X'(s) ds}\right]e^{k_{\mathrm{on}}\mu t},
\label{fxt2}
\end{eqnarray}
where $\mathrm{E}^{x+\mu}[...]$ denotes expectation over all possible sample paths of $X'$ over $[0,t]$, conditioned on $X'(0)=x+\mu$. Since $X'$ follows the CIR process Eq. (\ref{sde2}), the expectation $\mathrm{E}^{x+\mu}[...]$ has an expression similar to Eq. (5). In fact, the survival probability $f(x,t)$ in Eq. (\ref{fxt2}) equals
\[\left[\frac{\widetilde{\lambda}e^{\lambda t/2}/\sinh(\widetilde{\lambda}t/2)}
{\lambda+\widetilde{\lambda}\coth(\widetilde{\lambda}t/2)}\right]^{\alpha}
\exp\left[\frac{-2k_{\mathrm{on}}(x+\mu)}{\lambda+\widetilde{\lambda}\coth(\widetilde{\lambda}t/2)}+k_{\mathrm{on}}\mu t\right],
\]
with $\widetilde{\lambda}\equiv\sqrt{\lambda^2+2k_{\mathrm{on}}\lambda}$ and $\alpha = 4\mu$. At large $t$, this is
\begin{equation}
f(x,t)\sim \exp\left[-\frac{k_{\mathrm{on}}'(x+\mu)}{\lambda}-(2k_{\mathrm{on}}'-k_{\mathrm{on}})\mu t\right],
\label{asypf2}
\end{equation}
where $k_{\mathrm{on}}'\equiv 2k_{\mathrm{on}}/(1+\sqrt{1+2\theta'})$ and $\theta'\equiv k_{\mathrm{on}}/\lambda$. Thus, the dwell time distribution behaves asymptotically as
\begin{equation}
P(\widetilde{\tau}>t)\sim e^{-\widetilde{k}_{\mathrm{on}}\mu t}, \ \ \text{where} \ \ \ \widetilde{k}_{\mathrm{on}}=2k_{\mathrm{on}}'-k_{\mathrm{on}}.
\label{kon1}
\end{equation}
For $\theta'\ll1$, the asymptotic rate above becomes:
\begin{equation}
\widetilde{k}_{\mathrm{on}}=\left(\frac{3-\sqrt{1+2\theta'}}{1+\sqrt{1+2\theta'}}\right)k_{\mathrm{on}}\simeq \left(\frac{2-k_{\mathrm{on}}/\lambda}{2+k_{\mathrm{on}}/\lambda}\right)k_{\mathrm{on}}.
\label{kon2}
\end{equation}
The first equality above shows that $\widetilde{k}_{\mathrm{on}}$ can be negative when $\theta'>4$ or $k_{\mathrm{on}}>4\lambda$.
This arises as a curse of the possibility that $X(t)$ can go negative under Eq. (\ref{sde}).

When dealing with the Langevin approximation Eq. (\ref{lang}), researchers usually assume that the input noise is small enough (given a large copy number) such that the random variable $X$ could be replaced by its mean $\mu$ in Eq. (\ref{var}). This results in an OU approximation:
\begin{equation}
dX(t)=\lambda[\mu-X(t)]dt+\sqrt{2\nu}dW_t,
\label{ou}
\end{equation}
which takes a Gaussian distribution in steady state with $\sigma_X^2=\mu$. Compared
to the shifted Gamma distribution resulted from Eq. (\ref{sde}), the
Gaussian model of $X(t)$ has zero skewness and is unbounded below.
Thus, when $\mu$ is small, the OU approximation becomes an inappropriate
choice. By Feynman-Kac theorem, the survival probability $f(x,t)$ under Eq.
(\ref{ou}) satisfies
\begin{equation}
\frac{\partial f}{\partial t}=(\nu-\lambda x)\frac{\partial f}{\partial x}
+\nu \frac{\partial^2 f}{\partial x^2}-k_{\mathrm{on}}x f,
\label{pdef2}
\end{equation}
which can also be exactly solved. We omit the solution here, but later will show that $P(X(t)<0)>0$ for the OU process will lead to $\widetilde{k}_{\mathrm{on}}<0$ in certain regimes.

\begin{figure}
\scalebox{0.38}[0.38]{\includegraphics{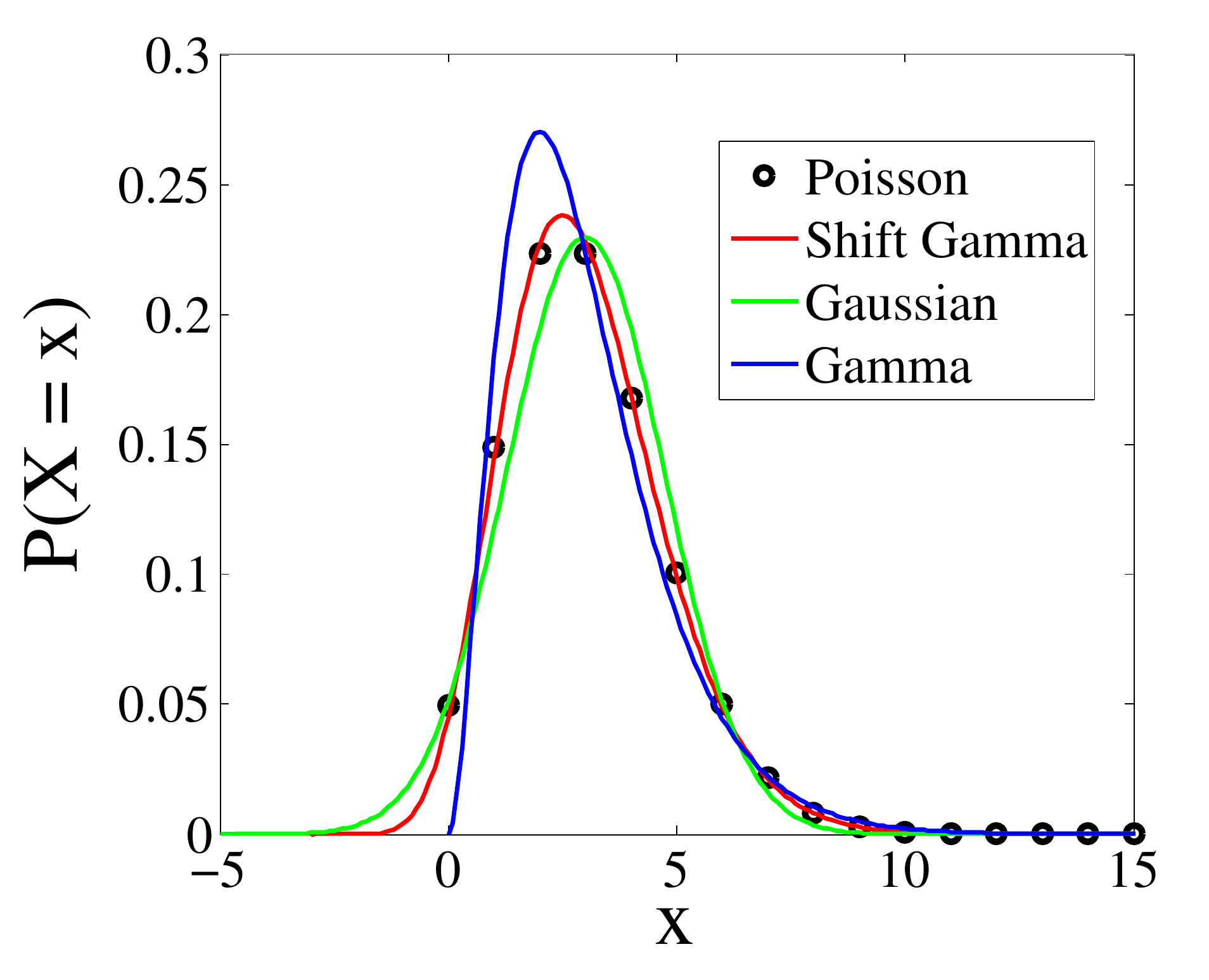}} \caption{(color online). Comparison of the Poisson, ``shifted" Gamma, Gaussian, and Gamma distributions, all of which have the same mean $\mu=3$ and the same variance $\sigma_X^2=\mu$.}
\end{figure}

We propose an alternative fix  for the Langevin approximation,  replacing the mean $\mu$ by its random counterpart $X(t)$ in Eq. (\ref{var}). This yields a CIR process:
\begin{equation}
dX(t)=\lambda[\mu-X(t)]dt+\sqrt{2\lambda X(t)}dW_t,
\label{cir2}
\end{equation}
under which $X(t)$ follows a Gamma distribution in steady state, with $\sigma_X^2=\mu$ as in all the previous diffusion approximations. The skewness in this model is found to be $2\mu^{-1/2}$, which is twice the skewness in the (birth-death) Poisson distribution. Though larger than desired, it is better than none (in the OU case). Fig. 4 plots a comparison of the Poisson, Gamma, shifted Gamma, and Gaussian distributions, all satisfying $\sigma_X^2=\mu$. One can see that the shifted Gamma distribution, which follows from Eq. (\ref{sde}), gives the closest approximation to the (birth-death) Poisson distribution, while the Gamma and Gaussian distributions are, roughly speaking, equally good to approximate the Poisson. Nonetheless, only the Gamma density from the CIR model is nonnegative, like the original Poisson distribution.

The diffusion approximations we have discussed so far, including Eq.
(\ref{cir}), Eq. (\ref{sde}), Eq. (\ref{ou}), and  Eq. (\ref{cir2}),
are all special cases of the following general It\'{o} SDE: \begin{equation}
dX(t)=\lambda[\mu-X(t)]dt+\sqrt{\sigma_0^2+\sigma_1^2 X(t)}dW_t.
\label{sde3}
\end{equation}
Under Eq. (\ref{sde3}) the survival probability $f(x,t)$ satisfies
\begin{equation}
\frac{\partial f}{\partial t}=\lambda(\mu- x)\frac{\partial f}{\partial x}
+\frac{1}{2}(\sigma_0^2+\sigma_1^2 x)\frac{\partial^2 f}{\partial x^2}-k_{\mathrm{on}}x f.
\label{pdef3}
\end{equation}
Again, a shortcut for solving Eq. (\ref{pdef3}) is to introduce $X'\equiv X+\sigma_0^2/\sigma_1^2$ which will evolve as a CIR process. This will allow us to make use of the existing results and obtain a similar expression for $f(x,t)$ as before. However, our main interest is the effective rate $\widetilde{k}_{\mathrm{on}}$ in the asymptotic behavior of $P(\widetilde{\tau}>t)\simeq \Omega\exp(-\widetilde{k}_{\mathrm{on}}\mu t)$, where $\Omega$ is some constant. Inspired by Eqs. (\ref{asypf}) and (\ref{asypf2}), we guess that when $t$ is sufficiently large, $f(x,t)\sim \exp(-Ax-\widetilde{k}_{\mathrm{on}}\mu t)$, for some constant coefficients $A$ and $\widetilde{k}_{\mathrm{on}}$ (to be determined). Plugging this expression into Eq. (\ref{pdef3}) yields:
\begin{equation}
-\widetilde{k}_{\mathrm{on}}\mu f=-A\lambda(\mu-x)f+\frac{1}{2}A^2(\sigma_0^2+\sigma_1^2 x)f-k_{\mathrm{on}}x f,
\end{equation}
which holds only when $A$ and $\widetilde{k}_{\mathrm{on}}$ jointly solve the following two algebraic equations:
\begin{eqnarray}
A\lambda\mu-\frac{1}{2}A^2\sigma_0^2 &=& \widetilde{k}_{\mathrm{on}}\mu,\\
A\lambda+\frac{1}{2}A^2\sigma_1^2&=&k_{\mathrm{on}}.
\end{eqnarray}

Given $\sigma_1^2>0$, the solution of Eq. (32) is:
\begin{equation}
A = \frac{-\lambda+\sqrt{\lambda^2+2k_{\mathrm{on}}\sigma_1^2}}{\sigma_1^2}
=\frac{2k_{\mathrm{on}}}{\lambda+\sqrt{\lambda^2+2k_{\mathrm{on}}\sigma_1^2}}.
\label{A}
\end{equation}
Thus, by defining $\theta'\equiv k_{\mathrm{on}}\sigma_1^2/\lambda^2$, Eq. (\ref{A}) becomes
\begin{equation}
A\lambda=\frac{2k_{\mathrm{on}}}{1+\sqrt{1+2\theta'}}\equiv k_{\mathrm{on}}'.
\end{equation}
Eliminating $A^2$ in Eqs. (31) and (32), we find that
\begin{equation}
\widetilde{k}_{\mathrm{on}}
=k_{\mathrm{on}}'\frac{\sigma_0^2+\mu\sigma_1^2}{\mu\sigma_1^2}-k_{\mathrm{on}}\frac{\sigma_0^2}{\mu\sigma_1^2}.
\label{kon3}
\end{equation} 
Clearly, when $\sigma_0^2=0$, Eq. (\ref{sde3}) becomes a CIR process and Eq. (\ref{kon3}) reduces to
$\widetilde{k}_{\mathrm{on}}=k_{\mathrm{on}}'=2k_{\mathrm{on}}/(1+\sqrt{1+2\theta'})$, recovering our result in Section III. When $\sigma_0^2/\sigma_1^2=\mu$, Eq. (\ref{kon3}) is $\widetilde{k}_{\mathrm{on}}=2k_{\mathrm{on}}'-k_{\mathrm{on}}$, coinciding with Eq. (\ref{kon1}).

Finally, if $\sigma_1^2=0$, the diffusion Eq. (\ref{sde3}) becomes an OU process and $A=k_{\mathrm{on}}/\lambda$ by Eq. (32). In this case,
\begin{equation}
\widetilde{k}_{\mathrm{on}}=k_{\mathrm{on}}\left(1-\frac{k_{\mathrm{on}}\sigma_0^2}{2\mu\lambda^2}\right)
=k_{\mathrm{on}}\left[1-\frac{k_{\mathrm{on}}\mu}{\lambda}\left(\frac{\sigma_X^2}{\mu^2}\right)\right],
\label{kon4}
\end{equation}
where $\sigma_X^2\equiv\sigma_0^2/(2\lambda)$ is the variance of the OU process. We can define $\theta$ in the same way as in Eq. (\ref{theta}), such that $\widetilde{k}_{\mathrm{on}}=k_{\mathrm{on}}(1-\theta/2)$ in Eq. (\ref{kon4}). This becomes negative if $\theta>2$, again a consequence of $P(X(t)<0)>0$. Note that the OU approximation Eq. (\ref{ou}) is a particular OU process with $\sigma_X^2=\mu$. This leads to $\widetilde{k}_{\mathrm{on}}=k_{\mathrm{on}}(1-k_{\mathrm{on}}/\lambda)$ in Eq. (\ref{kon4}). So for small $\theta$ or $\lambda\gg k_{\mathrm{on}}$, we have $\widetilde{k}_{\mathrm{on}}\simeq k_{\mathrm{on}}/(1+k_{\mathrm{on}}/\lambda)$, consistent with our result in \cite{Hu2011PRL}.

In sum, the general diffusion process defined by Eq. (\ref{sde3}) may take negative values with positive probability, unless $\sigma_0^2=0$ which corresponds to the CIR process. Negative biochemical input is unrealistic and may lead to unphysical results (such as $\widetilde{k}_{\mathrm{on}}<0$) if the input noise is large. For this reason, we conclude that the CIR process Eq. (\ref{cir}) is more suitable for modeling biochemical noise in the continuous setup. As shown, it is analytically tractable and possesses desirable statistical features, including stationarity, mean-reversion, Gamma distribution, and a tunable Fano factor.

\section{Conclusion}

In this paper, we have extended our previous research on the role of
noise in biological switching systems. We propose that a square-root
diffusion process can be a more reasonable model for biochemical
fluctuations than the commonly used OU process. We employ
standard tools in stochastic processes to solve a well-defined
fundamental biophysical problem. Consistent with our earlier results,
we find that the input noise acts to suppress the input-dependent
transitions of the switch. Our analytical results in this paper
indicate that this suppression increases with the input noise level as
well as the input correlation time. The statistical features uncovered
in this basic problem can provide us with new insights to understand
various experimental observations in gene regulation and signal
transduction systems. The current modeling framework may also be
generalized to incorporate other biological features such as
ultrasensitivity and feedbacks. Work along these lines is underway.

We would like to thank Ruth J. Williams, Yuhai Tu, Jose Onuchic,
and Wen Chen for stimulating discussions. This work has been
supported by the NSF-sponsored CTBP grant PHY-0822283.

\end{document}